\documentclass[%
 reprint,
superscriptaddress,
amsmath,amssymb,
 prb,
]{revtex4-2}

\usepackage{graphicx}
\usepackage{dcolumn}
\usepackage{bm}
\usepackage{float}
\usepackage{chemformula}
\usepackage{appendix, comment}

\usepackage[colorlinks=true,linkcolor=blue,anchorcolor=red,citecolor=blue,urlcolor=black]{hyperref}

\usepackage[mathlines]{lineno}

\newcommand{\RNum}[1]{\uppercase\expandafter{\romannumeral #1\relax}}

\begin{document}

\title{Interplay between Haldane and modified Haldane models in $\alpha$-$T_{3}$ lattice: \\Band structures, phase diagrams and edge states}

\author{Kok Wai Lee}
\affiliation{Science, Mathematics and Technology, Singapore University of Technology and Design, Singapore 487372, Singapore}

\author{Pei-Hao Fu}
\email{peihao\_fu@sutd.edu.sg}
\affiliation{Science, Mathematics and Technology, Singapore University of Technology and Design, Singapore 487372, Singapore}

\author{Yee Sin Ang}
\email{yeesin\_ang@sutd.edu.sg}
\affiliation{Science, Mathematics and Technology, Singapore University of Technology and Design, Singapore 487372, Singapore}

\begin{abstract}
We study the topological properties of the Haldane and modified Haldane models in $\alpha$-$T_{3}$ lattice. The band structures and phase diagrams of the system are investigated. Individually, each model undergoes a distinct phase transition: (i) the Haldane-only model experiences a topological phase transition from the Chern insulator ($\mathcal{C} = 1$) phase to the higher Chern insulator ($\mathcal{C} = 2$) phase; while (ii) the modified-Haldane-only model experiences a phase transition from the topological metal ($\mathcal{C} = 2$) phase to the higher Chern insulator ($\mathcal{C} = 2$) phase and we show that $\mathcal{C}$ is insufficient to characterize this system because $\mathcal{C}$ remains unchanged before and after the phase transition. By plotting the Chern number and $\mathcal{C}$ phase diagram, we show that in the presence of both Haldane and modified Haldane models in the $\alpha$-$T_{3}$ lattice, the interplay between the two models manifests three distinct topological phases, namely the $\mathcal{C} = 1$ Chern insulator (CI) phase, $\mathcal{C} = 2$ higher Chern insulator (HCI) phase and $\mathcal{C} = 2$ topological metal (TM) phase. These results are further supported by the $\alpha$-$T_{3}$ zigzag edge states calculations. Our work elucidates the rich phase evolution of Haldane and modified Haldane models as $\alpha$ varies continuously from $0$ to $1$ in an $\alpha$-$T_3$ model.

\end{abstract}
\maketitle

\section{INTRODUCTION}
Nontrivial topological states of matter in two-dimensional (2D) systems \cite{Hasan_Kane, Qi_Zhang} have garnered enormous research interest since the theoretical prediction of the quantum anomalous Hall insulator (QAHI) by the Haldane model  \cite{Haldane} and its later experimental observations \cite{Qi-Kun, Kim-Kee, Shao, Jotzu-Messer, Alba, Serlin_Tschirhart, Chen_Guorui, Sharpe_Fox}. QAHIs are also known as Chern insulators (CIs) because their topological phases are defined by an integer called the Chern number, $\mathcal{C}$ \cite{TKNN}. Originally, the QAHIs experimentally observed were only limited to $\mathcal{C} = 1$. Subsequently, QAHIs with $\mathcal{C} \geq 2$ were theoretically proposed \cite{Wang_Jing, Fang_Gilbert} and experimentally realized \cite{Zhao_Yi-Fan, Zhao_Zhang_Ruoxi}. Such states are termed higher Chern insulators (HCIs).   

Recently, it is demonstrated that the \emph{modified Haldane model} \cite{Colomes} can lead to 
antichiral edge states -- \emph{copropagating} edge states along the two parallel edges. The antichiral edge states are in stark contrast to the chiral edge states in QAHIs where the edge states are \emph{counterpropagating} along the two parallel edges. The antichiral edge states are achieved by modifying the Haldane mass term \cite{Haldane} so that it acts as a pseudoscalar potential to break the time-reversal symmetry and shift the energies of the two Dirac points in opposite directions. Alternatively, it has also been shown that the antichiral edge states can be realized via electron-phonon interaction \cite{Medina, Mella_José} and by combining two subsystems based on the original Haldane model with opposite chirality \cite{PhysRevB.104.L081401}. These edge states must be accompanied by counterpropagating gapless bulk states to ensure an equal number of left- and right-moving modes. As such, they can exist in topological metals (TMs), conducting materials with gapless band structures and localized edge states \cite{Cheng_Cerjan}. Various experimental platforms have been proposed \cite{Mandal, Wang_Chao_Zhang, Denner, Chen_Jianfeng, Bhowmick} to realize the antichiral edge states, and have been experimentally observed in a microwave-scale gyromagnetic photonic crystal \cite{Zhou_Peiheng}, topological circuit \cite{Yang_Yuting} and a 3D layer-stacked photonic metacrystal \cite{Liu_Jian-Wei}.

There is a strong interest in studying topological phases in different lattice structures, such as honeycomb \cite{Haldane, Kane_Mele_1, Ruegg}, Lieb \cite{Tsai_ChenFang, Goldman_Urban, Rui_Bin, Weeks_Franz, Zhu_Hou}, dice/$T_{3}$ \cite{Mondal_Sayan,Dey_Bashab_Kapri_Priyadarshini, 10.1063/1.5078627, PhysRevLett.81.5888, PhysRevB.64.155306, PhysRevB.103.155155, oriekhov2023size}, checkerboard \cite{Kai_Hong}, Kagom\'e \cite{Ohgushi, Xu_Gang, Guo_Franz, Liu_Zhu}, honeycomb Kagom\'e \cite{Zhang_Bingwen}, square \cite{Stanescu}, diamond \cite{Fu_Kane} and $\alpha$-$T_{3}$ lattices \cite{Wang_Liu, Bugaiko_2019}. The discovery of new topological phases in various lattices not only enriches the understanding of condensed matter physics, but also fuels potential technological applications \cite{gilbert2021topological, romeo2023experimental}. 
The $\alpha$-$T_{3}$ lattice \cite{Raoux} represents a particularly interesting lattice due to two prominent characteristics, namely a dispersionless zero-energy flat band and $\alpha$-dependent Berry phase which leads to interesting phenomena, such as super-Klein tunneling \cite{Betancur-Ocampo, Illes_Nicol} and unconventional quantum Hall effect \cite{PhysRevB.92.245410, PhysRevB.96.155301}. The $\alpha$-$T_{3}$ lattice is an extension of the graphene honeycomb lattice. In addition to the honeycomb A and B sites, an additional C site is introduced in the center of each hexagon, which coupled to either the A or B sublattice via the coupling strength, $\alpha t$. Here, $\alpha$ ($0 \le \alpha \le 1$) acts as a tuning parameter and $t$ is the A-B hopping term. As such, the $\alpha$-$T_{3}$ lattice serves as an interpolation between the graphene honeycomb ($\alpha = 0$) and dice/$T_{3}$ ($\alpha = 1$) lattices. Its low-energy dispersion consists of a Dirac cone and a dispersionless zero-energy flat band. At a critical doping, \ch{Hg_{1-$x$} Cd_{$x$}}Te can be mapped onto the $\alpha$-$T_{3}$ lattice with $\alpha = 1/\sqrt{3}$ \cite{Malcolm_Nicol}. $\alpha$-$T_{3}$ lattice can also be realized on optical platforms \cite{Raoux, Rizzi_Matteo}. Various aspects of the $\alpha$-$T_{3}$ lattice have been studied such as electro-magnetotransport properties \cite{Li_Fu_Zhang, Biswas, Islam_Dutta}, thermoelectric properties \cite{Xin_Pei_Wen, Alam_Waqas}, Andreev reflection \cite{Zhou_Xingfei}, Josephson effect \cite{Zhou_Xingfei}, Floquet engineering \cite{Dey_Bashab, Bashab_Ghosh, Tamang_Lakpa, PhysRevB.99.205135, PhysRevResearch.2.043245, PhysRevB.105.115309, PhysRevB.101.035129}, strain engineering \cite{Sun_Junsong} and the effect of Rashba spin-orbit coupling \cite{Lin_Fu}.

The Haldane model \cite{Haldane} in a honeycomb lattice gives rise to the Chern insulator phase while in a dice lattice, such model yields higher Chern insulator phases \cite{Mondal_Sayan, Dey_Bashab_Kapri_Priyadarshini, Filusch} with $\mathcal{C} = \pm 2$ \cite{Mondal_Sayan}. 
The $\alpha$-$T_{3}$ lattice, which interpolates between the honeycomb and the dice lattices, provides an interesting system to understand how the Haldane and modified Haldane terms affect the topological phases when $\alpha$ is tuned continuously from the `honecomb limit' at $\alpha = 0$ and the `dice limit' at $\alpha = 1$. 
How the two models jointly influence the topology of the $\alpha$-$T_{3}$ lattice when $\alpha$ is continuously tuned remains an open question. 

In this work, we study the possible topological phases in the $\alpha$-$T_{3}$ lattice that could emerge from the interplay between the Haldane and modified Haldane terms. We first demonstrate the topological properties of the individual cases by determining the Chern number, direct ($\Delta E_{Direct}$), and indirect ($\Delta E_{Indirect}$) band gaps. We argue that $\mathcal{C}$ is insufficient to characterize the modified Haldane model by showing that $\mathcal{C}$ remains unchanged before and after the system undergoes a phase transition. From the Chern number phase diagram and the $\alpha$-$T_{3}$ zigzag edge states, we show that the interplay between the two models in the $\alpha$-$T_{3}$ lattice manifests three distinct topological phases, namely the $\mathcal{C} = 1$ Chern insulator (CI) phase, $\mathcal{C} = 2$ higher Chern insulator (HCI) phase and $\mathcal{C} = 2$ TM phase. Our work elucidates the possible phases of the $\alpha$-$T_{3}$ lattice in the presence of the Haldane and modified Haldane terms, and shed light on the phase evolution of each model as $\alpha$ continuously varies from the honeycomb limit ($\alpha = 0$) to the dice limit ($\alpha = 1$). 

The remainder of this paper is organized as follows. In Sec. \ref{formalism}, the formulation is presented which includes the protocols of the Haldane and modified Haldane models in the $\alpha$-$T_{3}$ lattice, Hamiltonian and topological invariant. In Sec. \ref{results_discussion}, the results are presented which include the bulk band structures, phase diagrams and edge states. Lastly, in Sec. \ref{conclusion}, this paper is concluded with a brief summary of our results. 

\section{Model and FORMALISM}
\label{formalism}

\subsection{Model}
\label{model}

\begin{table}[tbp]
  \caption{Definitions of the nearest-neighbour (NN) and next-nearest-neighbour (NNN) vectors pointing from the B sites}
  \label{tab: Table 1}\renewcommand{\arraystretch}{1.55} 
  \begin{ruledtabular}
  \begin{tabular}{cc|cc}
  NN Vector  & Definition & NNN Vector & Definition  \\
  \colrule
  \bm{$\delta_{1}$} & $\left(\frac{\sqrt{3}}{2}a, \frac{1}{2}a\right)$ & \bm{$\delta_{1}'$} & $\left(\sqrt{3}a, 0\right)$ \\ 
  \bm{$\delta_{2}$} & $\left(0, a\right)$ & \bm{$\delta_{2}'$} & $\left(\frac{\sqrt{3}}{2}a, \frac{3}{2}a\right)$ \\
  \bm{$\delta_{3}$} & $\left(-\frac{\sqrt{3}}{2}a, \frac{1}{2}a\right)$ & \bm{$\delta_{3}'$} & $\left(-\frac{\sqrt{3}}{2}a, \frac{3}{2}a\right)$ \\ 
  \bm{$\delta_{4}$} & $\left(-\frac{\sqrt{3}}{2}a, -\frac{1}{2}a\right)$ & \bm{$\delta_{4}'$} & $\left(-\sqrt{3}a, 0\right)$  \\
  \bm{$\delta_{5}$} & $\left(0, -a\right)$ & \bm{$\delta_{5}'$} & $\left(-\frac{\sqrt{3}}{2}a, -\frac{3}{2}a\right)$  \\
  \bm{$\delta_{6}$} & $\left(\frac{\sqrt{3}}{2}a, -\frac{1}{2}a\right)$ & \bm{$\delta_{6}'$} & $\left(\frac{\sqrt{3}}{2}a, -\frac{3}{2}a\right)$ 
  \end{tabular}
  \end{ruledtabular}
\end{table}

\begin{figure*}[th]
    \centering  
    \includegraphics[width = \textwidth]{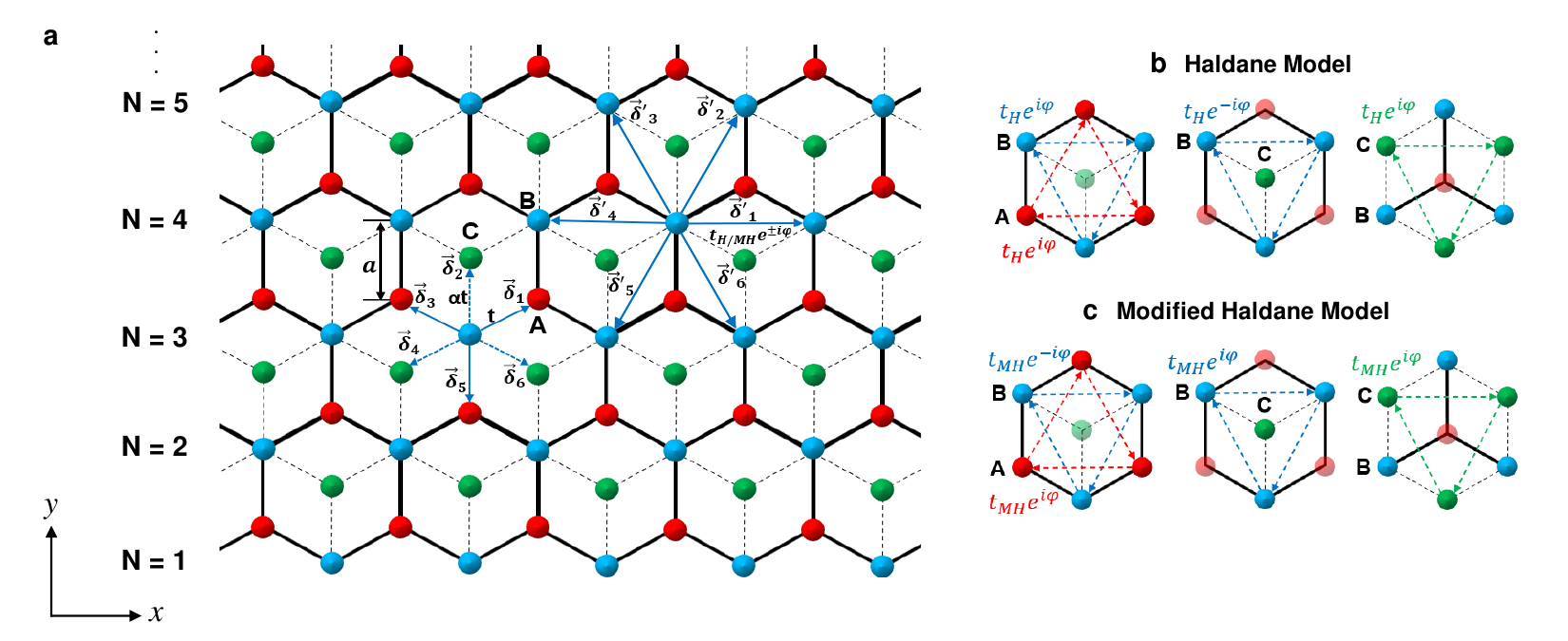}
    \caption{(a) Schematic of the $\protect\alpha $-$T_{3}$ lattice with zigzag edges. \bm{$\delta_{n}$} and  \bm{$\delta_{n}'$} ($n = 1$, $2$, $3$, $4$, $5$, $6$) denote the nearest-neighbour (NN) and next-nearest-neighbour (NNN) vectors pointing from the B sites respectively as defined in Table \ref{tab: Table 1}. Schematic of the (b)  Haldane and (c) modified Haldane NNN hoppings. The A, B and C sites are coloured in red, blue and green respectively.}
    \label{fig: Figure1}
\end{figure*}

The $\alpha$-$T_{3}$ lattice with both Haldane and modified Haldane terms as illustrated schematically in Fig. \ref{fig: Figure1} is described by the following Hamiltonian:
\begin{equation}
    \mathcal{H}=\mathcal{H}_{0}+\mathcal{H}_{H}+\mathcal{H}_{MH} \text{,}
    \label{real-space Hamiltonian}
\end{equation}
where the first term  
\begin{equation}
    \mathcal{H}_{0} = -\sum_{\langle ij\rangle }tc_{i}^{\dag }c_{j}-\sum_{\langle jk\rangle }\alpha tc_{j}^{\dag }c_{k} + H.c. \text{,}
\end{equation}%
describes the nearest-neighbour (NN) hoppings between the B and A (C) sites with strength $t$ ($\alpha t$). The second and third terms 
\begin{eqnarray}
    \mathcal{H}_{H} &=& \frac{t_{H}}{3\sqrt{3}}\left[ \sum_{\langle \langle
    ij\rangle \rangle }e^{-iv_{ij}\phi }c_{i}^{\dag }c_{j}+\alpha \sum_{\langle
    \langle jk\rangle \rangle }e^{-iv_{jk}\phi }c_{j}^{\dag }c_{k}\right] \nonumber \\
    && + H.c.
    \text{,}    
\end{eqnarray}
and 
\begin{eqnarray}
    \mathcal{H}_{MH} &=& \frac{t_{MH}}{3\sqrt{3}}\left[ \sum_{\langle \langle ij\rangle \rangle }e^{-i\mu _{i}v_{ij}\phi ^{\prime }}c_{i}^{\dag}c_{j} 
    + \alpha \sum_{\langle \langle jk\rangle \rangle }e^{-i\mu
    _{j}v_{jk}\phi ^{\prime }}c_{j}^{\dag }c_{k}\right] \nonumber \\
    && + H.c. \text{,}
\end{eqnarray}
are the Haldane and modified Haldane terms with strengths $t_{H}$ and $t_{MH}$ as well as phases $\phi$ and $\phi^{\prime}$ respectively. Here, $c_{i}^{\dag }$ ($c_{i}$) is the spinless fermionic creation
(annihilation) operator acting at the $i$th site, the summation of $\langle
ij\rangle $ ($\langle \langle ij\rangle \rangle $) runs over all the nearest (next-nearest)-neighbour sites, $H.c.$ denotes the  Hermitian conjugate, $v_{ij}=+1$ ($-1$) denotes the anticlockwise (clockwise) hopping and $\mu _{i}=+1$ ($-1$) denotes the next-nearest-neighbour (NNN) hoppings between A/C (B) sites. 

In contrast to the honeycomb lattice where an electron crosses from one sublattice to the other to hop to an NNN site (e.g., the path undertaken by an electron hopping from a B site to another B site is B-A-B), there are instead two possible paths for an electron to hop from a B site to another B site in the $\alpha -T_{3}$ lattice (i.e., B-A-B and B-C-B with hopping strengths $t_{H}$ and $\alpha t_{MH}$ respectively). All possible Haldane and modified Haldane NNN hopping paths are illustrated in Figs. \ref{fig: Figure1}(b) and (c) respectively.  

The resulting $k$-space Hamiltonian in the sublattice basis, $\left( |A_{k}\rangle,|B_{k}\rangle ,|C_{k}\rangle \right) ^{\top}$ obtained via the Fourier transformation of Eq. (\ref{real-space Hamiltonian}) is presented as follows: 
\begin{equation}
    \mathcal{\hat{H}}(\bm{k})=%
    \begin{bmatrix} G_{- -} && f^\dag\left(\bm{k}\right)
    && 0 \\ f\left(\bm{k}\right) && G_{+ -} + \alpha G_{- +} && \alpha
    f^\dag\left(\bm{k}\right) \\ 0 && \alpha f\left(\bm{k}\right) && \alpha
    G_{+ +}  \end{bmatrix}\text{,} 
    \label{k-space Hamiltonian}
\end{equation}
where
\begin{equation}
    f\left( \bm{k}\right) = -t\sum_{n = 1,3}^{5}e^{i\bm{k}\cdot \bm{\delta}_n} \text{,} 
\end{equation}
results from the conventional B-A NN hopping whereas 
\begin{equation}
    G_{\gamma \beta } = \frac{1}{3\sqrt{3}}\left[ t_{H}g\left( \bm{k},\gamma
    \phi \right) + t_{MH}g\left( \bm{k},\beta \phi ^{\prime }\right) \right] 
    \text{,}     
\end{equation}
\begin{equation}
    g\left(\bm{k}, \zeta\right) = \sum_{n = 1}^{6} e^{i[(-1)^{n}\zeta + \bm{k}\cdot \bm{\delta}_n^{\prime}]} \text{,}
\end{equation}
results from the Haldane and modified Haldane NNN hopping terms with $\zeta = \phi$ or $\phi'$, $\bm{k}=\left( k_{x},k_{y}\right) $ and indices $\gamma $, $\beta =\pm 1$.
The competition between the terms shall govern the possible phases of the system which are revealed directly by the bulk band structure obtained by solving the eigenvalue problem of Eq. (\ref{k-space Hamiltonian}) numerically. 

On the other hand, as to be demonstrated in Sec. \ref{chern}, the physics around the $\bm{K}^{\prime} = (-4\pi/3\sqrt{3}a, 0)$ and $\bm{K}=(4\pi/3\sqrt{3}a, 0)$ points is also focused on where the states are described by the following Dirac-like Hamiltonian: 
\begin{equation}
    \mathcal{H}_{\eta }\left( \bm{q}\right) =%
    \begin{bmatrix} L_{++}^{\eta} && \tilde{q} &&
    0 \\ \tilde{q}^\dag && L_{-+}^{\eta} + \alpha L_{+-}^{\eta} && \alpha\tilde{q} \\ 0 && \alpha
    \tilde{q}^\dag && \alpha L_{--}^{\eta} \end{bmatrix}\text{,}  
    \label{low-energy hamiltonian}
\end{equation}
where $\eta =+1$ ($-1$) represents the $K$ ($K^{\prime }$) valley with $%
\bm{q}=\left( q_{x},q_{y}\right) =\bm{k}-\bm{K}$ ($\bm{k}-\bm{K}^{\prime }$%
), $\tilde{q}= \hbar v_f\left( \eta q_{x}-iq_{y}\right) $, $\hbar v_f=3at/2$ and    
\begin{eqnarray}
L_{\gamma \beta }^{\eta} &=&\frac{t_{H}}{\sqrt{3}}\left[ -\cos \phi +\eta \sqrt{3}%
\sin \left( \gamma \phi \right) \right]   \nonumber \\
&&+\frac{t_{MH}}{\sqrt{3}}\left[ -\cos \phi ^{\prime }+\eta \sqrt{3}\sin
\left( \beta \phi ^{\prime }\right) \right] \text{,}
\end{eqnarray}%
serves as the Dirac mass term determining the bulk spectral gap.

The low-energy dispersion of Eq. (\ref{low-energy hamiltonian}) can be solved for analytically via the secular equation, $\det[\mathcal{H}_{\eta}( \bm{q}) - E_\eta ( \bm{q})] = 0$ which is, however, too complex to be presented in its full form here. For convenience, the three bands and their corresponding wavefunctions are labeled as $E_{m}^{\eta }( \bm{q})$ and $\psi _{m}^{\eta }$ respectively. The subscript, $m = -1$, $0$, and $+1$ denote the valence, middle, and conduction bands respectively. For our work, we let $\phi = \phi' = \pi/2$ to ensure the Haldane and modified Haldane NNN hoppings are purely imaginary \cite{Mondal_Sayan}. 

At the $K^{\prime}$ and $K$ points ($\tilde{q} = 0$), Eq. (\ref{low-energy hamiltonian}) becomes a diagonal matrix. Since a topological phase transition is usually related to a band gap closing-reopening process, we can define both the direct and indirect band gaps of our system in terms of the diagonal elements as follows:

\begin{subequations}
    \begin{equation}
    \Delta E_{Direct} = \vert (L_{-+}^{\eta} + \alpha L_{+-}^{\eta}) - \alpha L_{--}^{\eta}
    \vert
    \text{,} 
    \label{direct_band}
\end{equation}
\begin{equation}
    \Delta E_{Indirect} = \vert (L_{-+}^{+1} + \alpha L_{+-}^{+1}) - \alpha L_{--}^{-1}
    \vert
    \text{.} 
    \label{indirect_band}
\end{equation}
\end{subequations}

\subsection{Topological Invariant}
\label{chern} 
Typically, topological phases are associated with topological invariants which, for our system, is the Chern number, $\mathcal{C}$ of the valence band defined as follows: 
\begin{equation}
    \mathcal{C}=\mathcal{C}_{K} + \mathcal{C}_{K^{\prime }}\text{,}
\label{chern_valence}
\end{equation}
where $\mathcal{C}_{K (K^{\prime})}$ is the so-called valley Chern number defined as 
\begin{equation}
    \mathcal{C}_{\eta }=\frac{1}{2\pi }\iint_{BZ}\Omega _{\eta }\left( \bm{q}\right) \,d\bm{q} \text{,}
\end{equation}
and 
\begin{equation}
    \Omega _{\eta }(\bm{q})=i\text{Im}\sum_{m = 0,1}\frac{\langle \psi _{-1}^{\eta
    }|v_{x}|\psi _{m}^{\eta}\rangle \langle \psi _{m}^{\eta }|v_{y}|\psi
    _{-1}^{\eta}\rangle }{( E_{-1}^{\eta } - E_{m}^{\eta }) ^{2}}\text{,}
\end{equation}%
is the Berry curvature of the occupied band at the $\eta $-valley and $\bm{v} =\partial \mathcal{H}_{\eta}( \bm{q}) /\partial \bm{q}$. It is assumed that only the valence band ($m=-1$) is occupied.

\section{Results and Discussion}
\label{results_discussion}
Hereafter, the NN hopping strength, $t$ serves as the energy unit ($t = 1$) and the phases of the NNN hoppings, $\phi$ and  $\phi^{\prime}$ are fixed at $\pi/2$ \cite{Mondal_Sayan}. Each bulk band structure is plotted along the $k_{x}$ axis at $k_{y} = 0$, that is along the path joining the high-symmetry $K^{\prime}$, $M$ and $K$ points. The unit of $k_{x}$ is $k_{0} = 4\pi/\sqrt{3}a$ where $a$ is the graphene lattice constant taken to be $1$.

\subsection{Bulk Spectral and Topological Properties}
\label{bulk_spectral_topological}
\begin{figure*}[t]
    \centering  
    \includegraphics[width =
    \textwidth]{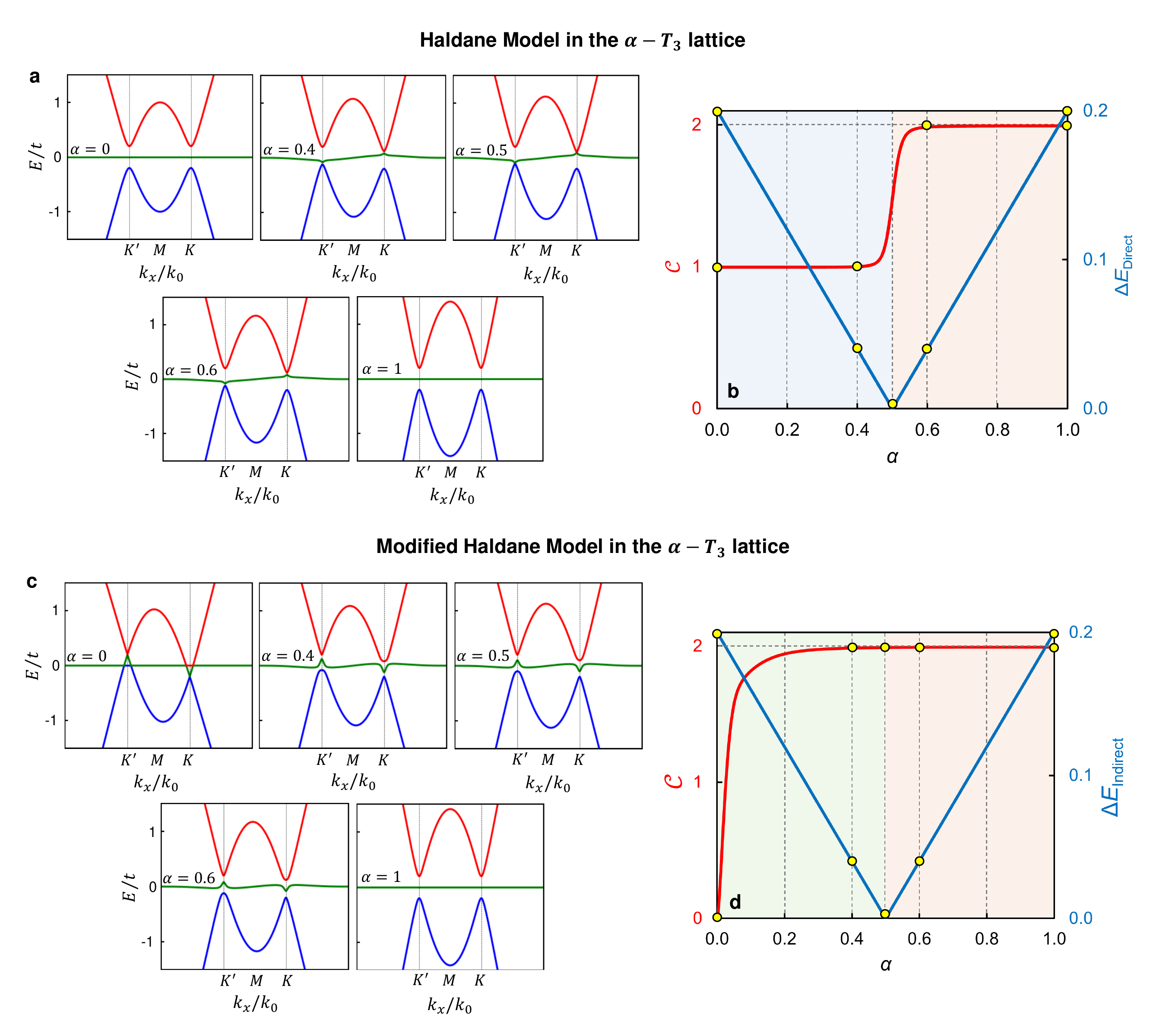}
    \caption{Evolution of the bulk band structures of the (a) Haldane and (c) modified Haldane models in the $%
    \protect\alpha$-$T_{3}$ lattice with respect to $\protect\alpha$. The conduction, middle and valence bands are coloured in red, green and blue respectively. (b) Chern number, $\mathcal{C}$ and direct band gap, $\Delta E_{\text{Direct}}$ vs $\protect\alpha$. 
    (d) Chern number, $\mathcal{C}$ and indirect band gap, $\Delta E_{\text{Indirect}}$ vs $\protect\alpha$. 
    The yellow dots denote the values of $\mathcal{C}$,  $\Delta E_{Direct}$ and $\Delta E_{Indirect}$  at $%
    \protect\alpha = 0$, $0.4$, $0.5$, $0.6$ and $1$. 
    For (a) and (b), $(t_{MH}, t_{H}) = (0, 0.2)t$ and $\phi = \phi^{\prime} = \pi/2$.  
    For (c) and (d), $(t_{MH}, t_{H}) = (0.2, 0)t$ and $\phi = \phi^{\prime} = \pi/2$.} 
    \label{separate_bulk_band_structure}
\end{figure*}

By setting the NNN hopping strengths, $(t_{MH}, t_{H}) = (0, 0.2)t$ and solving the eigenvalue problem of Eq. (\ref{k-space Hamiltonian}) numerically, we obtain the Haldane bulk band structure comprising three bands, namely the conduction, middle and valence bands. Figure \ref{separate_bulk_band_structure}(a) depicts its evolution with respect to $\alpha$ where it experiences a direct band gap opening-closing-reopening process. At $\alpha = 0$, akin to the graphene case, spectral gaps open at the $K^{\prime} $ and $K$ points due to the Haldane NNN hopping term, $t_{H}$ but the middle band remains flat owing to the presence of localized electrons at the C sites.
The middle band then becomes dispersive when $\alpha \neq 0$ due to the interaction between the B and C sublattices which causes the spectral gaps to shrink. As the value of $\alpha$  increases, the dispersive nature of the middle band becomes more prominent until it closes the spectral gaps at $\alpha = 0.5$. By further increasing $\alpha$ to $1$, the middle band returns to being dispersionless and the spectral gaps are recovered. 

The plot of the Chern number, $\mathcal{C}$ and direct band gap, $\Delta E_{Direct}$ against $\alpha$ is depicted in Fig. \ref{separate_bulk_band_structure}(b). Here, the blue and red regions represent the $\mathcal{C} = 1$ and $\mathcal{C} = 2$ topological phases respectively. Both a jump from $\mathcal{C} = 1$ (CI) to $\mathcal{C} = 2$ (HCI) and $\Delta E_{Direct} = 0$ at $\alpha = 0.5$ are observed, indicating that the topological phase transition corresponds to the closing of the direct band gap at $\alpha = 0.5$.

Similarly, by setting  $(t_{MH}, t_{H}) = (0.2, 0)t$, we obtain the modified Haldane bulk band structure. Figure \ref{separate_bulk_band_structure}(c) depicts its evolution with respect to $\alpha$ where it experiences an indirect band gap opening-closing-reopening process. At $\alpha = 0$, the bulk band structure is gapless and the band-touching points are shifted vertically in opposite directions due to the presence of the modified Haldane NNN hopping term, $t_{MH}$. The middle band becomes dispersive and spectral gaps open at the $K^{\prime}$ and $K$ points when $\alpha \neq 0$ as a result of the interaction between the B and C sublattices. In contrast to the Haldane model, the middle band becomes less dispersive as the value of $\alpha$ increases and it shrinks the indirect spectral gaps until they are closed at $\alpha = 0.5$. Again, the middle band returns to being dispersionless and the spectral gaps are recovered when $\alpha = 1$. 

The plot of the Chern number, $\mathcal{C}$ and indirect band gap, $\Delta E_{Indirect}$ against $\alpha$ is depicted in Fig. \ref {separate_bulk_band_structure}(d). Unlike the previous case, no jump in the value of $\mathcal{C}$ is observed. Instead, for infinitesimal values of $\alpha$ ($\alpha \rightarrow 0$) $\mathcal{C}$ is ill-defined due to the gapless spectrum. The sharp increase is not captured perfectly by Fig. \ref{separate_bulk_band_structure}(d) for want of computational accuracy. As $\alpha$ continues to increase, $\mathcal{C}$ attains a definite value of $2$. This shows that the Chern number is insufficient to characterize this particular system. On the other hand, $\Delta E_{Indirect}$ indeed becomes zero at $\alpha = 0.5$. Therefore, this system only experiences a phase transition from $\mathcal{C} = 2$ (TM) to $\mathcal{C} = 2$ (HCI) at $\alpha = 0.5$ as represented by the green and red regions of Fig. \ref{separate_bulk_band_structure}(d) respectively. Its topology remains unchanged. 

\begin{figure}[t]
    \centering 
    \includegraphics[width = 0.48\textwidth]
        {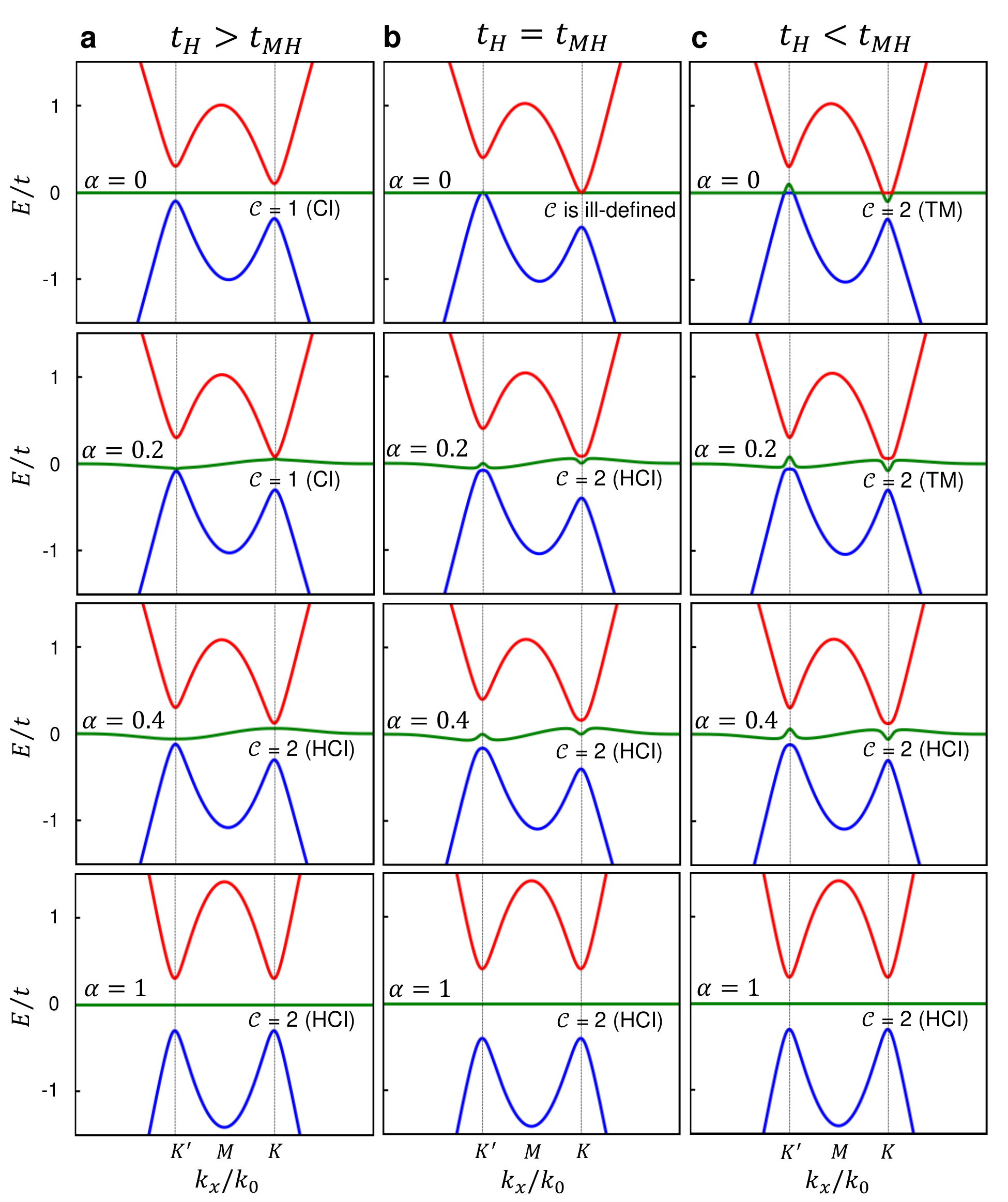}
    \caption{Evolution of the bulk band structure of the combined Haldane models in the $\protect\alpha$-$T_{3}$ lattice with
    respect to $\protect\alpha$ for (a) Case \uppercase\expandafter{\romannumeral 1\relax} - $t_{H} > t_{MH}$, (b) Case \uppercase\expandafter{\romannumeral 2\relax} - $t_{H} = t_{MH}$ and (c) Case \uppercase\expandafter{\romannumeral 3\relax} - $t_{H} < t_{MH}$. The values of the parameters are $(t_{MH}, t_{H}) = (0.1, 0.2)t$ 
    for Case \uppercase\expandafter{\romannumeral 1\relax}, $(t_{MH}, t_{H}) = (0.2, 0.2)t$ 
    for Case  \uppercase\expandafter{\romannumeral 2\relax}, $(t_{MH}, t_{H}) = (0.2, 0.1)t$ 
    for Case \uppercase\expandafter{\romannumeral 3\relax} and $\phi = \phi^{\prime}= \pi/2$ for all three cases. Each panel is labeled with its respective Chern number, $\mathcal{C}$. The conduction, middle and valence bands are coloured in red, green and blue respectively.}
\label{combined_haldane_models}
\end{figure}

Next, we consider three cases for the combined Haldane models in the $\alpha$-$T_{3}$ lattice: (i) Case \uppercase\expandafter{\romannumeral 1\relax} - $t_{H} > t_{MH}$, (ii) Case \uppercase\expandafter{\romannumeral 2\relax} - $t_{H} = t_{MH}$ and (iii) Case \uppercase\expandafter{\romannumeral 3\relax} - $t_{H} < t_{MH}$ where $(t_{MH}, t_{H}) = (0.1, 0.2)t$, $(0.2, 0.2)t$ and $(0.2, 0.1)t$ respectively. They are shown in Figs. \ref{combined_haldane_models}(a), (b) and (c) accordingly. In Case \RNum{1}, we obtain the spectral gaps at the $K^{\prime} $ and $K$ points due to $t_{H}$, and also their opposite vertical shifts due to $t_{MH}$ as shown in Fig. \ref{combined_haldane_models}(a) at $\alpha = 0$. Moreover, the presence of $t_{MH}$ leads to the system experiencing a topological phase transition from $\mathcal{C} = 1$ (CI) to $\mathcal{C} = 2$ (HCI) at $\alpha = 0.25$ instead of $0.5$ [Fig. \ref{separate_bulk_band_structure}(a)]. The opposite manner occurs for Case \RNum{3}. Here, we do not only obtain the shifts at the $K^{\prime}$ and $K$ points due to $t_{MH}$ but also the spectral gaps due to $t_{H}$ as shown in Fig. \ref{combined_haldane_models}(c) at $\alpha = 0$. Similarly, the presence of $t_{H}$ causes the system to experience a phase transition from $\mathcal{C} = 2$ (TM) to $\mathcal{C} = 2$ (HCI) at $\alpha = 0.25$ instead of $0.5$ [Fig. \ref{separate_bulk_band_structure}(c)]. On the other hand, as exemplified by Case \RNum{2} [Fig. \ref{combined_haldane_models}(b)], the system is  gapless (remains at $\mathcal{C} = 2$ (HCI)) at $\alpha = 0$ $(\alpha \neq 0)$ whenever $t_{H} = t_{MH}$. At $\alpha = 1$, Figs. \ref{combined_haldane_models}(a), (b) and (c) appear similar which can be explained by solving the eigenvalue problem of Eq. (\ref{low-energy hamiltonian}) as follows:
\begin{equation}
    \mathcal{H}^{\alpha = 1}_{\eta }\left( \bm{q}\right) =%
    \begin{bmatrix} \Delta && \tilde{q} &&
    0 \\ \tilde{q}^\dag && 0 && \tilde{q} \\ 0 && 
    \tilde{q}^\dag && -\Delta \end{bmatrix}\text{,}  
    \label{low-energy hamiltonian alpha_1}
\end{equation}
where  $\Delta= \eta (t_{H} + t_{MH})$ and the resulting eigenvalues are 
\begin{equation}
    E_0 = 0 \text{;} \hspace{0.4cm} 
    E_{\pm1}^{\eta} = \pm \sqrt{|\tilde{q}|^2 + \Delta }
    \text{.}
\end{equation}

\begin{figure}[tbp]
    \centering 
    \includegraphics[width =0.5\textwidth]
    {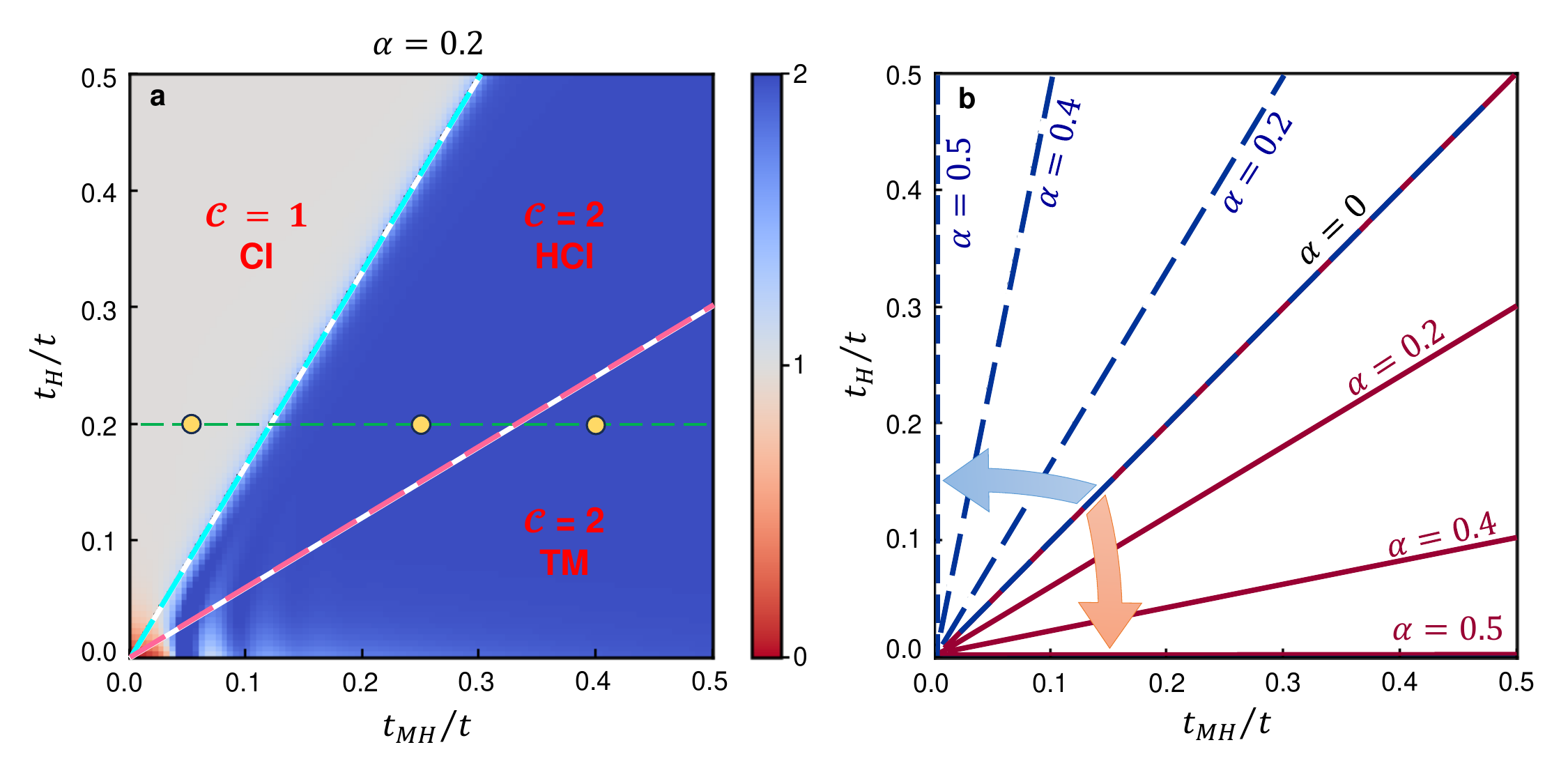}
    \caption{(a) Chern number, $\mathcal{C}$ phase diagram with respect to $t_{H} $ and $t_{MH}$ at $\protect\alpha = 0.2$ and $\protect\phi = \protect\phi^{\prime} = \protect\pi/2$. Here, CI, HCI and TM denote the terms Chern insulator, higher Chern insulator and topological metal respectively. The three yellow dots denote the values of $\mathcal{C}$ at $(t_{MH}, t_{H}) = (0.05, 0.2)t$, $(0.25, 0.2)t$ and $(0.4, 0.2)t$. (b) The variation of the CI-HCI (dashed blue) and HCI-TM (solid red) phase boundaries with respect to $\protect\alpha$ for $\protect\alpha = 0$, $0.2$, $0.4$ and $0.5$.}
    \label{fig: phase diagram}
\end{figure}

\begin{figure*}[tbp]
    \centering  
    \includegraphics[width =\textwidth]{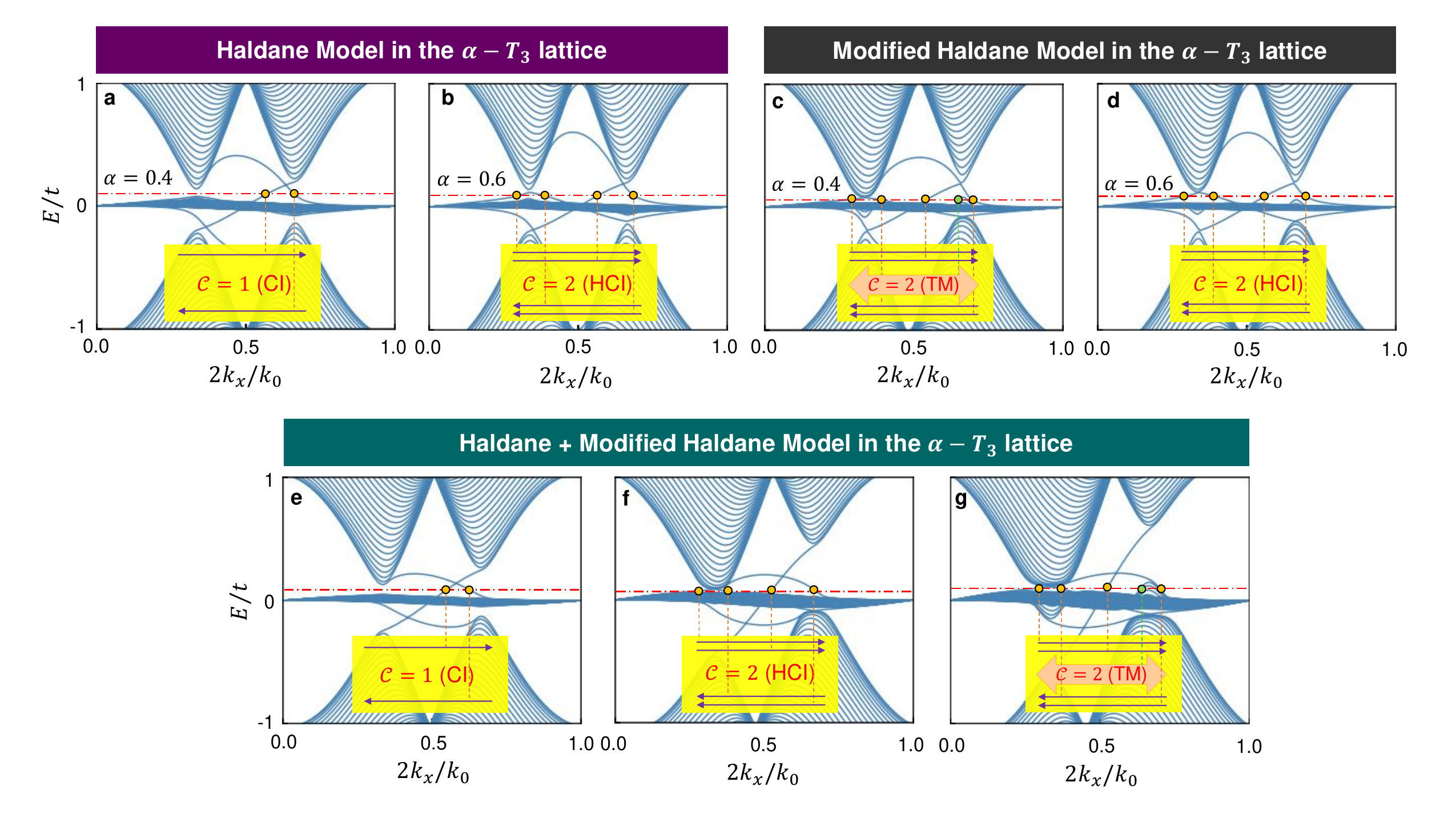}
    \caption{The band structure of the Haldane model applied to the $\protect\alpha$-$T_{3}$ zigzag nanoribbon at (a) $\protect\alpha = 0.4$ and (b) $\protect\alpha = 0.6$. The values of the parameters are $(t_{MH}, t_{H}) = (0, 0.2)t$ and $\phi = \phi^{\prime} = \pi/2$. 
    The band structure of the modified Haldane model applied to the $\protect\alpha$-$T_{3}$ zigzag nanoribbon at (c) $\protect\alpha = 0.4$ and (d) $\protect\alpha = 0.6$. The values of the parameters are $(t_{MH}, t_{H}) = (0.2, 0)t$ and $\phi = \phi^{\prime} = \pi/2$. (e) to (g) The band structures of the combined Haldane models applied to the $\protect\alpha$-$T_{3}$ zigzag nanoribbon corresponding to the three yellow dots in Fig. \protect\ref{fig: phase diagram}(a). The zigzag chain along the $y$-axis contains N = 50 AB sites.}
    \label{fig: edge states}
\end{figure*}

The Chern number, $\mathcal{C}$ phase diagram further demonstrates the interplay between $t_{H}$, $t_{MH}$ and $\alpha$ as depicted in Fig. \ref{fig: phase diagram}(a) for the specific case of $\alpha = 0.2$. Here, the combined Haldane models in the $\alpha$-$T_{3}$ lattice manifests a total of three phases, namely the $\mathcal{C} = 1$ (CI) phase, $\mathcal{C} = 2$ (HCI) phase and $\mathcal{C} = 2$ (TM) phase. For instance, at $t_{H} = 0.2t$, increasing $t_{MH}$ from $0.05t$ to $0.4t$ [going from the left to right in Fig. \ref{fig: phase diagram}(a)] causes the system to first experience the CI phase, followed by the HCI phase and finally the TM phase. The CI-HCI and HCI-TM phase boundaries are determined by the closing of the direct and indirect band gaps respectively. Figure \ref{fig: phase diagram}(a) exhibits fluctuations at the lower left region, indicating $\mathcal{C}$ is ill-defined due to the system being gapless.

The variation of the CI-HCI and HCI-TM phase boundaries with respect to $\alpha$ is depicted in Fig. \ref{fig: phase diagram}(b) which satisfy the following relations:
\begin{subequations}
    \begin{equation}
    t_{H} = \frac{t_{MH}}{1 - 2\alpha} \text{,}
    \label{direct_equation}
\end{equation}
\begin{equation}
    t_{H} = (1 - 2\alpha)t_{MH} \text{,}
    \label{indirect_equation}
\end{equation}
\end{subequations}
respectively. The relations are derived by equating Eqs. (\ref{direct_band}) and (\ref{indirect_band}) to zero. As a result, at $\alpha = 0$, the phase boundaries are degenerate, restoring the graphene case \cite{Colomes}. As $\alpha$ increases, the slopes of the CI-HCI and HCI-TM phase boundaries increases and decreases respectively, opening the $\mathcal{C} = 2$ HCI phase regime which eventually dominates the entire phase diagram when $\alpha \geq 0.5$.

\subsection{Edge States}
\label{edge_states}
The concept of the bulk-edge correspondence (BEC) states that topological phases possess localized edge states protected by non-trivial bulk topological invariants \cite{Hatsugai, Hatsugai_2, Halperin}. 
Therefore, the evolutions of the Chern numbers as well as the phases correspond to that of the edge states of the system.

To plot the edge states, the band structure of the $\alpha$-$T_{3}$ zigzag nanoribbon \cite{Fujita, Nakada} is obtained by considering open boundary condition along the $y$-direction and periodic boundary condition along the $x$-direction with zigzag edges. 
The sites along the $y$-direction are labelled as A\textsubscript{1}, B\textsubscript{1}, C\textsubscript{1}, A\textsubscript{2}, B\textsubscript{2}, C\textsubscript{2},..., A\textsubscript{N}, B\textsubscript{N}, C\textsubscript{N}, etc. 
A schematic of the $\alpha$-$T_{3}$ nanoribbon is illustrated in Fig. \ref{fig: Figure1}(a). 

Figure \ref{fig: edge states} depicts the crossings of the zigzag edge states with the Fermi level. For the case of the Haldane model in the $\alpha$-$T_{3}$ lattice, initially, there is one edge state in each edge propagating in opposite directions [Fig. \ref{fig: edge states}(a)] which is consistent with $\mathcal{C} = 1$ (one chiral edge state), signifying the Chern insulator (CI) phase. After the critical point ($\alpha = 0.5$), there are two edge states in each edge propagating in opposite directions [Fig. \ref{fig: edge states}(b)] which is consistent with $\mathcal{C} = 2$ (two chiral edge states), signifying the higher Chern insulator (HCI) phase.

For the case of the modified Haldane model in the $\alpha$-$T_{3}$ lattice, before $\alpha = 0.5$, there are two edge states in each edge propagating in opposite directions accompanied by bulk states [Fig. \ref{fig: edge states}(c)], indicating the topological metal (TM) phase. $\mathcal{C} = 2$ (two chiral edge states) does not encode information regarding the bulk states. After $\alpha = 0.5$, the Fermi level does not cross the bulk states and only the two chiral edge states remain ($\mathcal{C} = 2$) [Fig. \ref{fig: edge states}(d)], indicating the higher Chern insulator (HCI) phase. 

Figures \ref{fig: edge states}(e) to (g) visualize the topological phases manifested by the combined Haldane models as the $\mathcal{C}$ phase diagram of Fig. \ref{fig: phase diagram}(a) is traversed horizontally to the right. At $(0.05, 0.2)t$, we first obtain one edge state in each edge propagating in opposite directions corresponding to $\mathcal{C} = 1$ (one chiral edge state) [Fig. \ref{fig: edge states}(e)], indicating the CI phase. Next, at $(0.25, 0.2)t$, we obtain two edge states in each edge propagating in opposite directions corresponding to $\mathcal{C} = 2$ (two chiral edge states) [Fig. \ref{fig: edge states}(f)], indicating the HCI phase. Finally, at $(0.4, 0.2)t$, we obtain two chiral edge states ($\mathcal{C} = 2$) and bulk states, indicating the TM phase. 

\section{CONCLUSION}

\label{conclusion} 
In summary, we study the topological properties of the
Haldane and modified Haldane models in the $\alpha$-$T_{3}$ lattice, both individually and collectively. Firstly, we demonstrate that each model manifests a distinct phase transition. The Haldane model experiences a topological phase transition at $\alpha = 0.5$ from the Chern insulator ($\mathcal{C} = 1$) phase to the higher Chern insulator ($\mathcal{C} = 2$) phase. For the modified Haldane model, it experiences a phase transition from the topological metal ($\mathcal{C} = 2$) phase to the higher Chern insulator ($\mathcal{C} = 2$) phase at $\alpha = 0.5$. 
The fact that $\mathcal{C}$ remains $2$ indicates that the Chern number is insufficient to characterize the modified Haldane model. From the Chern number $\mathcal{C}$ phase diagram, we show that the interaction between the Haldane and modified Haldane parameters realizes three distinct topological phases, namely the $\mathcal{C} = 1$ Chern insulator (CI) phase, $\mathcal{C} = 2$ higher Chern insulator (HCI) phase and $\mathcal{C} = 2$ topological metal (TM) phase. Furthermore, we investigate how the tuning parameter, $\alpha$ influences the phases. At $\alpha = 0$, the system only has the CI and TM phase regimes which is the graphene case. As $\alpha$ increases, the HCI phase regime is created and dominates the entire phase diagram at $\alpha \geq 0.5$. The Chern numbers and phases of the aforementioned cases are corroborated by plotting the zigzag edge states. 
Finally, we remark that we can include more effects into our current model such as the intrinsic \cite{Wang_Liu} and Rashba \cite{Lin_Fu} spin-orbit couplings (SOCs), Floquet engineering \cite{Dey_Bashab, Tamang_Lakpa, Qin_Fang1, Qin_Fang2, PhysRevB.99.205135, PhysRevResearch.2.043245, PhysRevB.105.115309, PhysRevB.101.035129} and strain engineering \cite{Sun_Junsong} in order to potentially realize new possible topological phases.

\begin{acknowledgments}
This work is supported by the Singapore Ministry of Education (MOE) Academic Research Fund (AcRF) Tier 2 Grant (MOE-T2EP50221-001 9).
\end{acknowledgments}

\providecommand{\noopsort}[1]{}\providecommand{\singleletter}[1]{#1}%

\end{document}